\documentclass[12pt]{iopart}
\usepackage{iopams}
\usepackage{graphicx}

\begin{document}
\title[Microscopic Fission Dynamics]{Microscopic Description of Nuclear Fission Dynamics}

\author{A S Umar$^1$, V E Oberacker$^1$, J A Maruhn$^2$ and P-G Reinhard$^3$}
\address{$^1$Department of Physics and Astronomy, Vanderbilt University, Nashville, Tennessee 37235, USA}
\address{$^2$Institut f\"ur Theoretische Physik, Goethe-Universit\"at, D-60438 Frankfurt am Main, Germany}
\address{$^3$Institut f\"ur Theoretische Physik, Universit\"at Erlangen, D-91054 Erlangen, Germany}
\ead{umar@compsci.cas.vanderbilt.edu}

\begin{abstract}
We discuss possible avenues to study fission dynamics starting from a time-dependent mean-field
approach.
Previous attempts to study fission dynamics using the time-dependent Hartree-Fock (TDHF)
theory are analyzed.
We argue that different initial conditions may be needed to describe fission dynamics depending
on the specifics of the fission phenomenon and propose various approaches towards this goal.
In particular, we provide preliminary calculations for studying fission following a heavy-ion
reaction using TDHF with a density contraint. Regarding prompt muon-induced fission,
we also suggest a new approach for combining the time-evolution of the muonic wave function
with a microscopic treatment of fission dynamics via TDHF.
\end{abstract}
\submitto{\textbf{Open Problems in Nuclear Structure Theory}, \JPG}


\section{Introduction}
A fully microscopic description of the nuclear fission process is one of the most
challenging problems in theoretical
nuclear physics.  Attacking this problem in terms of determinantal many-body wavefunctions,
as is commonly done,
entails the understanding of single-particle evolution, proper coupling to all
of the relevant collective degrees of freedom, and incorporating the proper quantum mechanical framework.
In principle, the semi-classical limit of this process would involve the summation over an ensemble of quantum
mechanical fission paths that would explain the experimentally observed mass distributions.
The problem is further exacerbated by the rich selection of initial conditions that lead to fission.
In addition to most
commonly studied forms of fission, namely spontaneous or neutron induced fission,
which to a large degree may be influenced by the zero temperature potential energy surface,
the fission events that occur during a heavy-ion collision, e.g.
quasi-fission and fusion-fission or compound
nucleus fission, involve excited composite systems that may not be fully equilibriated.
For example, it has recently been shown that as we increase the temperature of the system the potential energy surfaces are
completely changed in comparison to the $T=0$ case~\cite{PN09}.

The conclusion one can draw from the above discussion is that it is
desirable to develop a dynamical approach to fission. 
Microscopic theories for fission have
been discussed in the past in the context of quantized
adiabatic time-dependent Hartree-Fock theory (ATDHF)~\cite{GRR83} and
in terms of the path-integral method~\cite{Ne82,GRR82}. 
These approaches have  been developed around the concept of a
collective fission path and are biased on sub-barrier processes,
mostly spontaneous fission or perhaps low-energy fusion. A bit more
extended dynamics of fission in a collective model space has been
studied via the time-dependent generator coordinate method employing a
Hartree-Fock-Bogoliubov (HFB) basis with quadrupole and octupole
constraints~\cite{GB05}.  
While static properties influencing fission, in particular the
structure of the collective potential energy surface, are customarily
studied via adiabatic microscopic mean-field
theories~\cite{DG08,Sch09},
the description of excited fission channels in the 
classically allowed regime above the barrier requires truly dynamical
approaches which are not confined to a collective subspace.  
The method of choice is here the TDHF theory. It becomes
particularly suited for fission studies when
coupled with a novel method of using a
density-constraint to obtain the corresponding TDHF trajectory on the
multi-dimensional potential energy surface of the combined nuclear
system~\cite{CR85,US85}.  Here, TDHF provides the evolution of the
nuclear shape (density) including all of the self-consistent dynamical
effects present in the mean-field limit.  The density-constrained TDHF
(DC-TDHF) approach has been successfully used to calculate dynamical
ion-ion interaction barriers~\cite{UO06b,UO06d,UO07a,UO08a,UO09a} for
fusion reactions.  In addition, one-body energy dissipation extracted
from TDHF for low-energy fusion reactions was found to be in agreement
with the friction coefficients based on the linear response theory as
well as those in models where the dissipation was specifically
adjusted to describe experiments~\cite{WL09}.  However, while the
initial conditions for a heavy-ion reaction are well defined, the same
is not true for fission.  Various types of fission reactions are
characterized by different initial conditions which may not be obvious
at the onset.

In this manuscript we will first discuss previous attempts to study fission via the standard
TDHF theory. This may provide clues to the choice of initial conditions. We then perform
preliminary studies of fission using the DC-TDHF approach coupled with a collective boost.
Finally, we discuss a different approach to study fission dynamics, namely prompt muon-induced
fission involving nuclear dynamics described by TDHF.

\section{Previous TDHF Studies and Initial Conditions}
In general there are very few attempts to generalize the static mean-field approach
to a dynamical approach for fission.
In this section we will discuss some of the previous attempts to study fission using the
standard TDHF approach.

Perhaps the most well known dynamical fission study is the one discussed in Ref.~\cite{NK78}.
Here, the authors studied the fission of $^{236}$U using TDHF with axial and reflection
symmetries and with a simplified form of the Skyrme interaction without the spin-orbit force.
The initial state was determined via a constrained Hartree-Fock (CHF) calculation to be
$1$~MeV beyond the saddle point. Furthermore, due to the imposed axial symmetry restriction
a time-dependent pairing had to be introduced in order to force the coupling of otherwise unmixed
states. This is an important point to stress because the significance of pairing for the
general fission problem is still an open question. Here, pairing was solely introduced to
break unphysical symmetries. As anticipated, the authors do find a number of fission paths that
strongly depend on the choice of the pairing properties. This seems to suggest that to study
dynamics of fission no symmetry restrictions should be made. However, this study does not
directly address the problem of how the system evolves to the chosen initial state and
naturally suffers from too many approximations made to simplify the numerical calculations.

Another early study of fission via TDHF is given in Ref.~\cite{OI83}.
Here, the authors perform a similar study except the initial CHF states are chosen from
single-center or two-center determinantal trial functions and no pairing. These states
are then boosted with a collective quadrupole velocity field. One observes that
the two-center initial states can be induced to fission while the single-center
calculations do not result in fission despite the deposition of large amount of 
collective energy. The authors draw the conclusion that the problem lies with having a single Slater
determinant and the lack of understanding as to which degree of freedom to invoke for the
correct initialization of fission dynamics.

Finally, an earlier TDHF study limited to slabs of nuclear matter~\cite{DN81} and using
a velocity field to trigger fission leads to the observation that starting from the
HF ground-state does not produce fission for reasonable velocity fields, while starting
from HF states that are obtained by lifting some nucleons from occupied states to previously
unoccupied levels fission can be invoked with relative ease.

The conclusion one can draw from all these studies is that the initial many-body state
for fission should be some form of a correlated state that contains couplings to continuum
or other excited states. The actual construction of this initial state will also depend
on the type of fission that one is planning to study, prompt fission being perhaps the
most difficult one since here the correlations and virtual excitations are the
sole drivers to fission. For the case of neutron induced fission, the selection of
where the neutron energy is deposited in terms of the single-particle picture and
the construction of the resulting correlated state is required.
The case of fusion-fission and especially quasi-fission can be better understood in
terms of the states formed during a TDHF collision since these do contain multiple
centers and large couplings to the continuum. This was noted by a statement attributed to
A.~K.~Kerman in Ref.~\cite{DD81} as: \textit{...it is tempting to speculate that states
formed in TDHF fusion reactions might be associated with single-particle wave functions
which are approximate eigenstates of the instantaneous HF Hamiltonian and the many-body
wave function constructed from these single-particle wave functions could then be
considered a transition state to processes that are not taken into account in TDHF
theory}.
The problem these authors faced was the unavailability of a numerical method to extract
the undissipated energy from TDHF determinantal wave functions since evolving
to infinite time was not an option. They tried the imaginary-time method to find the
corresponding static solutions but the numerical procedure always lead to the ground
state of the combined system. This now can be accomplished using the density constraint
combined with the imaginary-time method, as it used in DC-TDHF calculations and will be
discussed in the next section.

\section{TDHF with Density Constraint}
In order to validate some of the above findings we have tried to achieve dynamical
fission by starting from either the ground state or the traditional fission isomer
obtained by CHF calculations with a quadrupole constraint. Specifically, we have
looked at $^{240}$Pu and $^{238}$U systems.
The difference from earlier calculations was that here fully three-dimensional
TDHF codes with no symmetry assumptions and the full modern Skyrme forces
were used.
Using collective boost operators of the type $e^{\pm \imath p q_{L0}(\mathbf{r})}$
we were unable to obtain fission for reasonable excitation energies.
If a large amount of collective energy is deposited we observe symmetric fission,
which for the case of $^{240}$Pu is the least probable outcome.
We are still investigating this approach by using a combination of quadrupole and
octupole boost operators in an attempt to channel the system to the appropriate
fission path. However, it is becoming apparent that the energy may have to
be deposited in a more selective way than to the entire nucleus. We are also
studying the possibility of constructing excited many-body states and using
them as an initial state for fission.
Naturally, all of these studies only apply to the case of neutron induced fission.

\begin{figure}[!htb]
\begin{center}
\includegraphics*[height=.21\textheight]{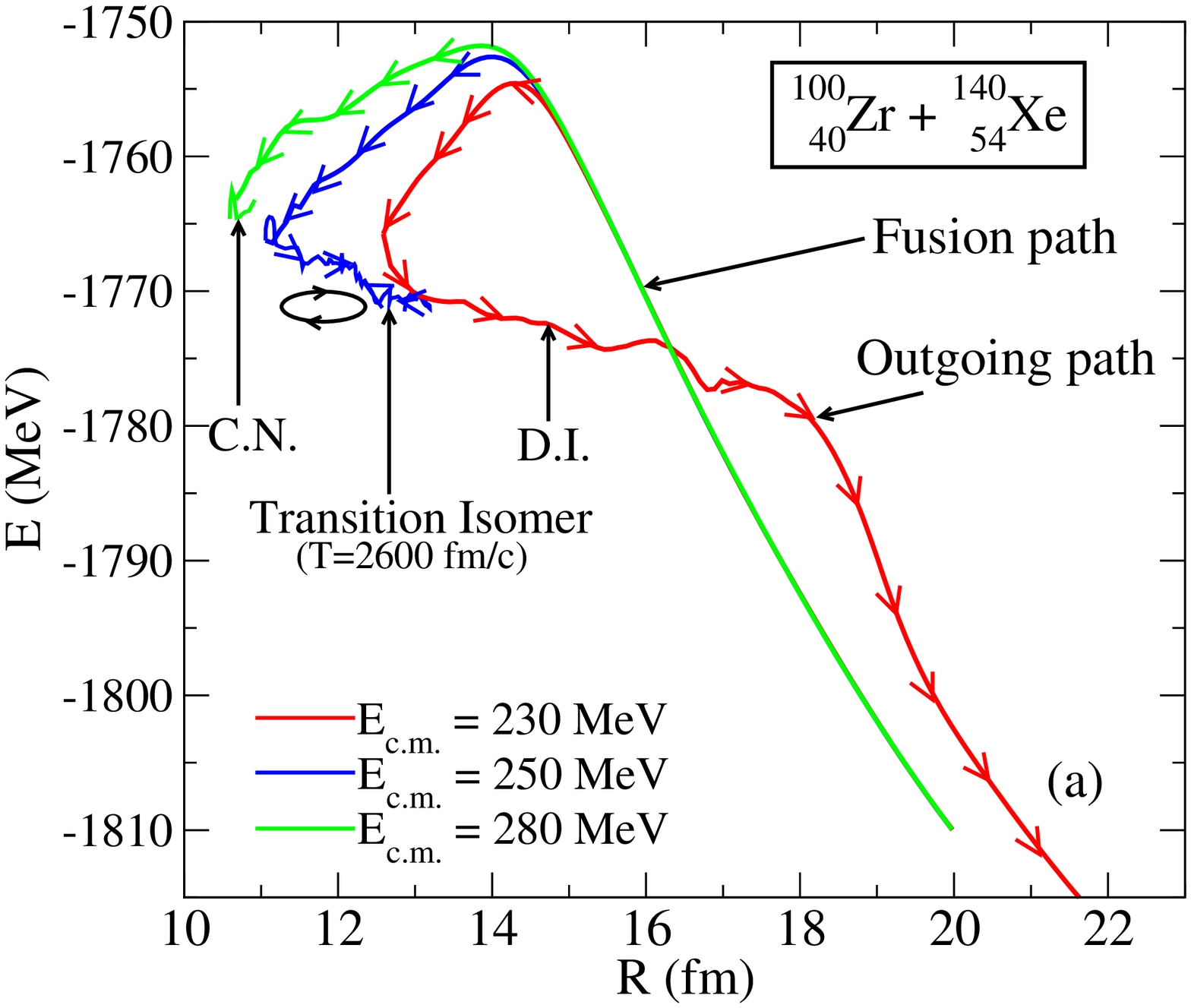}\includegraphics*[height=.21\textheight]{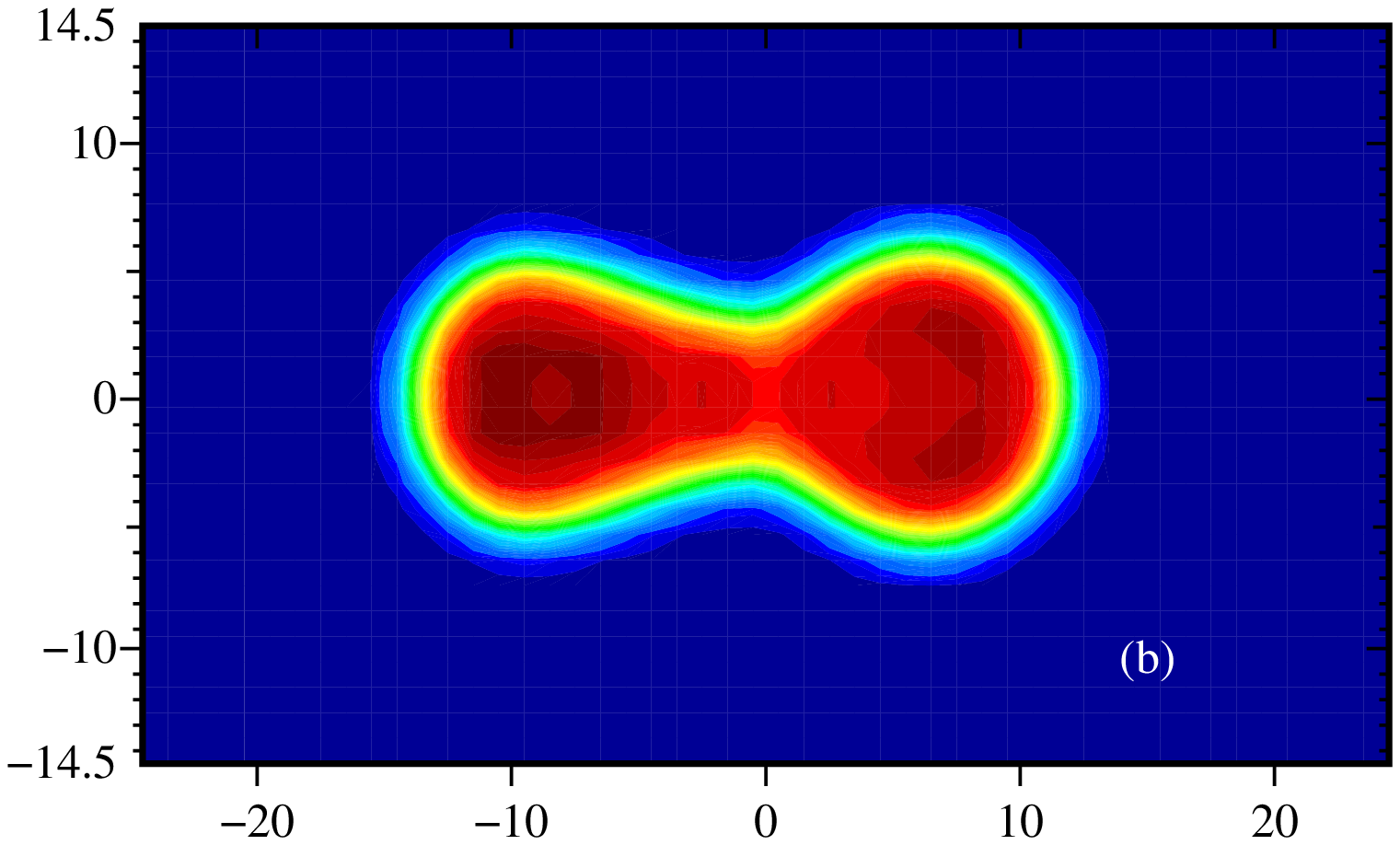}
\caption{\label{fig1}\\(a) Ion-ion potential for the $^{100}$Zr+$^{140}$Xe system at three different energies.
\\ (b) New shape isomer of $^{240}$Pu found from the transition state.}
\end{center}
\end{figure}

\begin{figure}[!htb]
\begin{center}
\includegraphics*[height=.26\textheight]{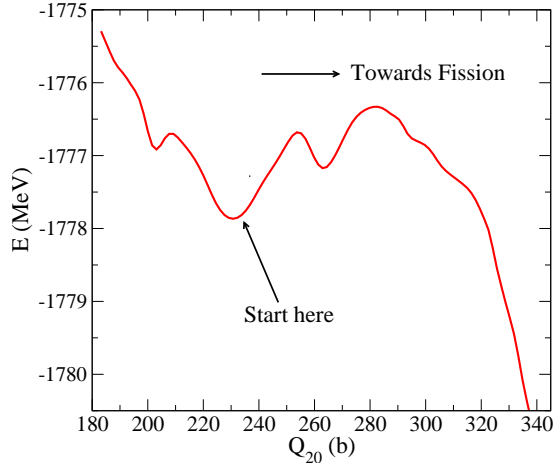}
\caption{\label{fig2} Potential barrier in the vicinity of the new shape isomer.}
\end{center}
\end{figure}

As we have  also discussed above, fission following a heavy-ion collision
(quasi-fission or fusion-fission) may be studied by creating the initial states
via the actual TDHF evolution of the collision followed by density constraint
calculation.
We have first performed collisions of $^{100}$Zr+$^{140}$Xe,
which results in the extensively studied $^{240}$Pu composite system. 
The choice of these nuclei were motivated by the most probable fragment mass and
charge for the induced  fission of $^{240}$Pu.
Along the TDHF
trajectory the density-constraint was applied to obtain the ion-ion potentials. In Fig.~\ref{fig1}a
we plot the result at three different bombarding energies.
As we see at the lowest energy the system undergoes a deep-inelastic collision, while at the
highest energy we observe complete fusion, as seen from the fact that the system is void of
any collective motion for a long time (2500~fm/c). On the other hand, for the intermediate
energy the system executes large collective oscillations with very little damping for the same
time interval. This is what we call the \textit{transition state} and indicates that the system is
trapped in an isomeric minimum and executes collective motion. Following this idea we have
started from a density along this path and performed an \textit{unconstrained} minimization.
The result was an isomer of the  $^{240}$Pu system shown in Fig~\ref{fig1}b ($\beta_2=2.27$,
$Q_{20}=230b$, and $Q_{30}=-28b^{3/2}$).
This isomer is far from
the well known isomer of $^{240}$Pu, which cannot be reached in such reactions. One feature of these
transition states is that they fission with a very small collective boost, e.g. $e^{\pm \imath p q_{20}(\mathbf{r})}$.
Using $p=0.0025$ the nucleus fissions with an initial excitation energy of $7$~MeV.
By using the above boost and performing
DC-TDHF calculations for the fissioning nuclei one can investigate the potential barriers in the vicinity
of the isomeric minimum. This is shown in Fig.~\ref{fig2}.  The resulting masses of the two fragments
are $A_1,Z_1=106,42$ and $A_2,Z_2=134,52$.
We feel that this approach to study fission following a heavy-ion reaction is very promising and
further explorations will be done in the near future.

\section{Fission ``clocks''}
In this section we will discuss a different approach to study fission
dynamics, namely muon-induced fission, which may also shed light on to
nuclear energy dissipation defined as the irreversible flow of energy
from collective to intrinsic single-particle motion~\cite{Ha78}.
There are two different mechanisms that contribute to nuclear energy
dissipation: two-body collisions and one-body friction. The latter
mechanism is caused by the moving \textit{walls} of the  one-body time-dependent
mean field as described by TDHF. The role played by these two
dissipation mechanisms in fission and in heavy-ion reactions is not yet
completely understood. Assuming that friction is caused by
two-body collisions, Davies et al.~\cite{DSN76} calculated the effect of
viscosity on the dynamics of fission; they extracted a viscosity coefficient
$\mu = 0.015$ Tera Poise from a comparison of theoretical and experimental
values for the kinetic energies of fission fragments. The corresponding time
delay for the nuclear motion from the saddle to the scission point was found
to be of the order of $\Delta t =1 \times 10^{-21}$ s. However, in one-body
dissipation models the time delay is an order of magnitude larger.

Several experimental techniques are sensitive to the energy dissipation
in nuclear fission. At high excitation energy, the multiplicity of
pre-scission neutrons \cite{Ga87} or photons \cite{Ho95} depends on the
dissipation strength. At low excitation energy, the process of
prompt muon-induced fission~\cite{WJ78,MO80,ON80,RB91,OU93,O99} provides
a suitable \textit{clock}.

Following the formation of an excited muonic atom, inner shell
transitions may proceed by inverse internal conversion where the 
muonic excitation energy is transferred to the nucleus. In actinides,
the $2p \rightarrow 1s$ and the $3d \rightarrow 1s$ muonic
transitions result in excitation of the nuclear giant dipole and giant
quadrupole resonances, respectively, which act as 
doorway states for fission. The nuclear excitation energy is typically
$6.5 - 10$ MeV, i.e. $E^*$ exceeds the fission barrier. Because the muon lifetime is long compared
to the timescale of prompt nuclear fission, the motion of the muon in
the Coulomb field of the fissioning nucleus may be utilized to learn
about the dynamics of fission. If there is large friction between the
outer fission barrier and the scission point the muon will
remain in the lowest molecular energy level and emerge in the $1s$ bound
state of the heavy fission fragment. On the other hand, if friction is
small, there is a non-zero probability that the muon may be promoted
to higher-lying molecular orbitals from where
it will end up attached to the light fission fragment. Therefore,
theoretical studies of the muon-attachment probability to the light
fission fragment, $P_L$, in combination with experimental data can be
utilized to analyze the dynamics of fission.

From a theoretical point of view, prompt muon-induced fission has several
attractive features. Because $E^*$ exceeds the fission barrier we do not
face the difficult problem of a microscopic description of barrier
tunneling. The muon dynamics is determined by solving either the
Schr\"odinger equation~\cite{MO80,ON80} or the Dirac equation~\cite{OU93,O99} 
in the presence of a time-dependent external Coulomb field which is
generated by the nuclear charge density during fission. So far, the motion
of the fissioning nucleus has been described by a simple classical model
involving a linear friction force.

Our time-dependent Dirac equation calculations~\cite{O99} predict a strong mass asymmetry
dependence of the muon attachment probability $P_L$ to the light fission
fragment, in agreement with experimental data. The theory also predicts
a dependence of $P_L$ on the dissipated energy. By
comparing our theoretical results to the experimental data of
ref. \cite{RB91} we extract a dissipated energy of order $0-10$ MeV for
$^{237}$Np. The $10$ MeV value corresponds to a fission time delay from
saddle to scission of order $2 \times 10^{-21}$ s.
This value of $E_{\rm diss}=10$ MeV
agrees with results from other low-energy fission
measurements that are based on the odd-even effect in the charge yields
of fission fragments \cite{Wa91}. 

In the future, we might be able to test the validity of one-body
(mean-field) energy dissipation in large-amplitude collective motion
by combining the time-evolution of the muonic wave function (as described by
the 3-D Dirac equation) with a microscopic treatment of fission
dynamics with our 3-D TDHF code.

\section{Summary}

The study of fission dynamics represents one of the most challenging problems
in nuclear many-body theory.
In the absence of a full microscopic and quantum mechanical theory for the many-body
tunneling process we ask the question about the generalization of the mean-field
based calculations, that have been extensively used to calculate potential
energy surfaces and fission barriers, to the dynamical case using a TDHF based
approach.
A careful analysis leads to the conclusion that this may be possible provided
the proper initial conditions for the dynamical calculations are chosen.
We show that the most straightforward study using TDHF may be fission which follows a
heavy-ion reaction.
This could be achieved via the use of density constraint and collective boost
operators on top of the TDHF dynamics.
Studying the dynamics of fission induced by low energy neutrons requires the
construction of the initial state which represents the state of the nucleus
after the energy transfer. One could construct such excited states
by using standard variational procedures, which we are currently investigating.
Clearly, all of these initial states will contain correlations beyond the
simple CHF state. Exactly what type of correlations are needed could emerge
from such studies.

\section*{Acknowledgments}
This work has been supported by the U.S. Department of Energy under grant No.
DE-FG02-96ER40963 with Vanderbilt University, and by the German BMBF
under contract Nos. 06FY159D and 06ER142D.

\end{document}